# Yukarı Yönlü Dikgen Olmayan Çoklu Erişimde Derin Öğrenme Sezicileri ile Pilot Aralığı Azaltılması

# Pilot Interval Reduction by Deep Learning Based Detectors in Uplink NOMA

*Ahmet EMİR, Ferdi KARA, Hakan KAYA*
Elektrik Elektronik Mühendisliği
Zonguldak Bülent Ecevit Üniversitesi
Zonguldak, Türkiye
{ahmet.emir,f.kara,hakan.kaya}@beun.edu.tr

*Özetçe*—Dikgen Olmayan Çoklu Erişim (Non-Orthogonal Multiple Access-NOMA) tekniğinin dikgen çoklu erişim (orthogonal multiple Access-OMA) tekniklerinden spektral verimliliği fazladır. Kanalın alıcıda bilinmediği yukarı yönlü haberleşme sistemlerinde farklı zaman dilimlerinde her bir kullanıcıdan gönderilen pilot sinyaller NOMA'nın spektral verimliliğini azaltmaktadır. Bu çalışmada yukarı yönlü haberleşme sistemlerinde, baz istasyonunda kullanıcılardan gönderilen pilot sinyallerin yanıtının bilindiği derin öğrenme (DL-deep learning) temelli seziciler araştırılmıştır. Kullanıcılar için tek pilot gönderilerek DL sezicilerdeki zaman adımı azaltılarak NOMA'nın spektral verimliliğinin korunması hedeflenmiştir.

*Anahtar Kelimeler — Dikgen Olmayan Çoklu Erişim(NOMA); derin öğrenme, pilot aralığı azaltımı*

*Abstract*— Non-Orthogonal Multiple Access (NOMA) has higher spectral efficiency than orthogonal multiple access (OMA) techniques. In uplink communication systems that the channel is not known at the receiver, pilot signals sent from each user in different time intervals have reduced the spectral efficiency of NOMA. In this study, in the uplink communication system, DL-deep learning based detectors which are known to respond to the pilot signals sent from the users at the base station have been researched. It is aimed to maintain the spectral efficiency of NOMA by sending a single pilot from users, thus reducing the time interval in the DL detectors.

I. Giriş

Dikgen olmayan çoklu erişim (Non-orthogonal multiple Access -NOMA) spektral kazanç açısından kazançsağlamaktadır ve bu sayede çoklu erişim teknikleri içerisinde 5G için aday gösterilmektedir [1,2]. Literatürde NOMA ile ilgili çalışmalar çoğunlukla kapasite ve kesinti olasılığı ile ilgilidir [3-5]. Dikgen çoklu erişim (Orthogonal Multiple Access-OMA) sistemleri ile karşılaştırıldığında kapasite avantajına rağmen NOMA sistemleri hata olasılığı açısından dezavantajlıdır. Hata performansının artırılması için ardışık girişim giderici (successive interference canceler -SIC) tabanlı seziciler yerine Derin Öğrenme (Deep Learning-DL) tabanlı seziciler önerilmiştir [6-8]. Diğer taraftan SIC tabanlı seziciler ya da DL tabanlı sezicilerde alıcı tarafta kanal kestirimi yapabilmek için pilot sinyallere ihtiyaç duymaktadır. Yukarı yönlü NOMA'da her bir kullanıcının pilot sinyali farklı bir zaman diliminde (time interval) Baz İstasyonuna (Base Station- BS)'e gönderilmektedir [9]. Bu durumda bir NOMA sembolü gönderebilmek için kullanıcılardan gelen pilot sinyaller ve NOMA verisi olmak üzere toplam üç zaman dilimine ihtiyaç duyulmaktadır. Bu durumda spektral verim azalmaktadır ve NOMA'nın sağladığı avantaj ortadan kaybolmaktadır.

Bu çalışmada spektral verim açısından bu üç zaman dilimli DL tabanlı sezicilerin getirdiği dezavantajın yok edilmesi için her iki kullanıcının pilot sinyalinin tek zaman diliminde *BS*'e gönderildiği yukarı yönlü (uplink)-NOMA sistem modeli önerilmiştir. Böylelikle NOMA sembolü ve pilot sinyal toplamda iki zaman diliminde *BS*'e ulaşmaktadır. Önerilen sistem modelinde kanal kestirimi ve sinyal sezimini ayrı ayrı yapmak yerine DL tabanlı bir ortak sinyal sezim işlemi yapılmaktadır. Alınan pilot ve veri sinyallerine dayanarak, *BS* kanal bilgisine sahip olmadan kullanıcıların sinyallerini sezebilmektedir. Ayrıca, karşılaştırma yapabilmek için önerilen iki zaman dilimli DL seziciye ek olarak literatürde tanımlanmış üç zaman dilimli DL sezici de tasarlanmıştır. Bu sistem modelinde iletişim *BS*'e gönderilen NOMA sinyaline ek olarak $UE_1$ (Gezgin Aygıt –User Equipment)'den gelen pilot sinyal ve $UE_2$'den gelen pilot sinyal nedeniyle üç zaman diliminde gerçekleşmektedir. İki zamanlı DL sezicide ise ilk zaman diliminde yakın ve uzak kullanıcılara ait pilot sinyaller süperpozisyon kodlu (superposition coded –SC) olarak tek zaman diliminde *BS*'e gönderilmiştir, ikinci zaman diliminde ise NOMA sembolleri *BS*'e gönderilmiştir. Gönderilen pilot sembol sayısı arttıkça SIC tabanlı sezicilerden iki zaman dilimli DL sezicinin üstünlüğü gösterilmiştir.

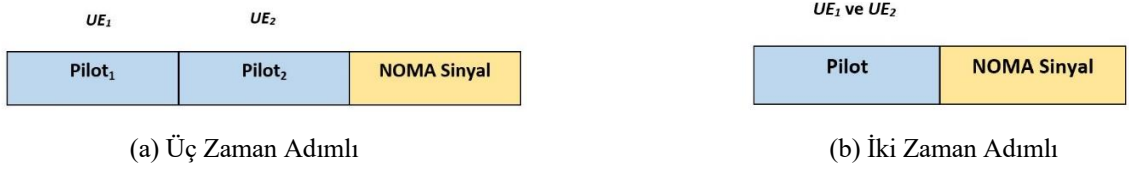

(a) Üç Zaman Adımlı                   (b) İki Zaman Adımlı

Şekil 1  Pilot Yaklaşımlı Yukarı Yönlü NOMA sistem modelleri

Çalışmanın bundan sonraki bölümleri şu şekilde sunulmuştur. Bölüm II'de Geleneksel Yukarı Yönlü-NOMA sistem modeli ve önerilen DL tabanlı yukarı yönlü-NOMA sezme işlemi tanıtılmıştır. Bölüm III'te DL için benzetim sonuçları sunulmuştur. Son olarak Bölüm IV'te sonuçlar tartışılarak çalışma sonlandırılmıştır.

## II. SİSTEM MODELİ

### A. Geleneksel Yukarı Yönlü NOMA

Bu çalışmada iki mobil kullanıcı ($UE_1$ ve $UE_2$) ve bir baz istasyonundan ($BS$) oluşan yukarı yönlü NOMA iletişim modeli sunulmuştur. $UE_1$ baz istasyonuna $UE_2$ 'den daha yakındır. Bu nedenle $UE_2$'nin kanal koşulları $UE_1$'den daha kötüdür. $UE_1$ yakın kullanıcı $UE_2$ uzak kullanıcı olarak isimlendirilmiştir. Bütün düğümlerin tek antenli olduğu düşünülmüştür. $BS$'e gelen süperpozisyon kodlu (superposition coded –SC) sembol:

$$y = \sqrt{P_1}x_1 h_1 + \sqrt{P_2}x_2 h_2 + n \qquad (1)$$

ile gösterilir.

Burada $P_1$ $UE_1$'in iletim gücünü, $P_2$ $UE_2$'nin iletim gücünü göstermektedir. $x_1$ ve $x_2$ sırasıyla $UE_1$ ve $UE_2$ için temel bant sayısal modülasyonlu karmaşık sembolleri temsil etmektedir. $UE_1$ ve $UE_2$ için sırasıyla Dikgen Faz Kaydırmalı Anahtarlama (Quadrature Phase Shift Keying -QPSK) ve İkili Faz Kaydırmalı Anahtarlama (Binary Phase Shift Keying -BPSK) kullanıldığı varsayılmıştır. $h_1$ $UE_1$ ile BS arasındaki $\sigma_1^2$ varyanslı sönümlemeli kanal katsayısını, $h_2$ ise $UE_2$ ile BS arasındaki $\sigma_2^2$ varyanslı sönümlemeli kanal katsayısını ifade etmektedir. $n$ ise $BS$'de oluşan güç spektral yoğunluğu $N_0/2$ olan toplanır beyaz Gauss gürültüsüdür.

Alıcıda kanal durum bilgisi bilinmediği durumda pilot sinyal gönderme yöntemleri ile sinyal sezimi yapılması gerekmektedir. Bu durumda yukarı yönlü NOMA sistemlerinde Şekil1(a)'da görüldüğü üzere birinci zaman diliminde $UE_1$ 'den gelen pilot sinyali ikinci zaman diliminde $UE_2$ 'den gelen pilot sinyali üçüncü zaman diliminde NOMA sinyali $BS$'e ulaşmaktadır. Denklem (2)'de her bir pilot zaman adımında $BS$'e ulaşan pilot sembol verilmiştir.

$$y_{i,p} = \sqrt{P_i}x_{p,i}h_i + n \quad i = 1,2 \qquad (2)$$

Burada $x_{p,i}$ $i$. $UE$'den $BS$ 'e gönderilen pilot sinyali ifade eder. İlk olarak alıcıya gelen bu pilot sinyalleri kullanılarak kanal kestirimi yapıldıktan sonra $BS$'de en büyük olabilirlikli (Maximum Likehood-ML) sezici ile doğrudan yakın kullanıcı sinyallerine karar verilir. Daha sonra yakın kullanıcı sinyali gelen sinyalden SIC işlemi ile çıkartılır. Sonrasında ML sezici ile uzak kullanıcı sembolleri kestirilir.

Geleneksel Yukarı Yönlü NOMA sistemlerinde pilot yaklaşımlı sinyal kestirimleri üç zaman adımlıdır. Üç zaman adımlı sinyal sezicide yakın kullanıcıda ulaşılabilir hız (achievable rate) aşağıdaki gibidir [5]:

$$R_1 = \frac{1}{3}\log_2(1 + SINR_1) = \frac{1}{3}\log_2\left(\frac{P_1|h_1|^2}{P_2|h_2|^2+N_0}\right) \qquad (3)$$

Burada SINR kullanıcının İşaret Girişim Artı Gürültü Oranını (Signal to Interference Plus Noise Ratio) ifade etmektedir.

Üç zaman adımlı sinyal sezicide uzak kullanıcıdaki ulaşabilir hız ise Denklem (4)'teki gibidir [5]:

$$R_2 = \frac{1}{3}\log_2(1 + SINR_2) = \frac{1}{3}\log_2\left(\frac{P_2|h_2|^2}{N_0}\right) \qquad (4)$$

### B. Önerilen Yukarı Yönlü NOMA DL Sistem Modeli

Üç zaman adımlı Yukarı Yönlü NOMA sistemlerinde kullanıcılardan gelen pilot sinyaller ayrı zaman dilimlerinde $BS$'e gönderilmektedir Bunun yerine önerilen sistem modelinde uzak ve yakın kullanıcılardan gelen pilotlar toplanarak tek bir pilot sinyal $BS$'e gönderilmektedir. Toplanarak $BS$'e gönderilen $UE_1$ ve $UE_2$'ye ait pilot sinyallerin toplamı aşağıdaki gibi verilmektedir.

$$y_p = \sqrt{P_1}x_{p,1}h_1 + \sqrt{P_2}x_{p,2}h_2 + n \qquad (5)$$

Şekil 1(b)'den görüleceği üzere önerilen sistemde bir zaman diliminde NOMA sinyalleri diğer zaman diliminde pilot sinyaller $BS$'e ulaşmaktadır. Önerilen DL ağı, Şekil 2'deki gibi beş katmandan oluşmaktadır: giriş katmanı, iki seri bağlı Uzun kısa süreli bellek (Long short term memory-LSTM) katmanı, Tam Bağlı (Fully connected-FC) katmanı ve Regresyon (REG) katmanı.  Giriş katmanına, Denklem (1)'de verilen NOMA sembolü ve Denklem (5)'te verilen pilot sembolünün fazör (I- In Phase) ve dikgen (Q-Quadrature) bileşenleri bir çerçeve olarak verilmektedir ($y^{(I)}, y^{(Q)} - y_p^{(I)}, y_p^{(Q)}$).  Bir çerçeve uzunluğu toplam $N$ adet sembolden oluşmaktadır. Giriş katmanı, sinyali LSTM (Long Short Term Memory- Uzun Dönem Kısa Hafıza) katmanlarına aktarmaktadır. Birinci ve ikinci LSTM katmanları, sırasıyla 80 ve 60 gizli hücreden oluşur.

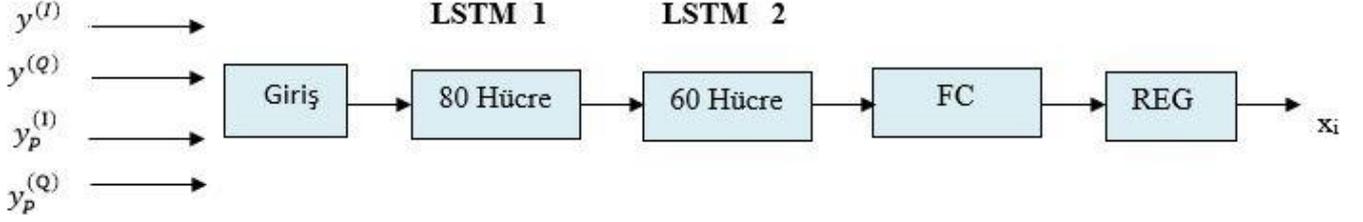

Şekil 2 DL tabanlı Sezici Sistem Modeli

LSTM pilot sinyal dizisi ve giriş sinyal dizisi arasındaki uzun süreli bağımlılıkları öğrenme kabiliyetinden dolayı tercih edilmiştir. LSTM katmanı, önceki durumu ezberleyerek ve her zaman adımında geçerli duruma ekleyerek yinelemeyi destekler. .Bir LSTM hücresinin her bir bileşeninde, giriş ağırlıkları ve yinelenen (recurrent) ağırlıklar öğrenilmiş ve güncellenmiştir. Tam bağlı (Fully connected-FC) katmanı, ikinci LSTM katmanından gelen sinyali bir ağırlık matrisi ile çarpar ve daha sonra buna bir son (bias) vektörü ekler. REG (Değer Atama-Regresyon) çıkış katmanı $UE_i$ 'nin karmaşık modüle edilmiş sinyalini kestirir. REG katmanının performansı aşağıda verilen Yarı ortalama kare hata kaybı (Half mean squared error loss-HMSE) fonksiyonu ile değerlendirilir.

$$E = \frac{\sum_{j=1}^{N}(x_i(j)-\widehat{x_i}(j))^2}{N} \quad (6)$$

Oluşturulan model, benzetim verileri ile Tablo 1'de verilen eğitim parametreleri için çevrimdışı olarak eğitilmiştir.

TABLO I  DL SEZİCİ EĞİTİM PARAMETRELERİ

| NOMA KULLANICI SAYISI | 2 |
|---|---|
| KANAL MODELİ | DURAĞIMSI RAYLEIGH KANALI |
| KULLANILAN EĞİTİM PROGRAMI | MATLAB DEEP LEARNING TOOLBOX |
| KULLANILAN DL MODELİ | LSTM |
| KULLANILAN PİLOT YÖNTEMLERİ | İKİ ZAMAN ADIMLI,ÜÇ ZAMAN ADIMLI |
| YIĞIN (BATCH) BOYUTU | 10000 |
| EĞİTİM ÖRNEĞİ SAYISI | $10^7$ |
| PİLOT/ÇERÇEVE ORANI | 1/2 |
| ÖĞRENME ORANI | 0.01 |
| LSTM HÜCRE SAYILARI | 80,60 |
| ÇERÇEVE SAYILARI | 8,16,32,64,128 |
| PİLOT/ÇERÇEVE ORANI | 1/2 |
| KAYIP FONKSİYONU | HMSE |
| OPTİMİZASYON | ADAM |

İki zaman adımlı DL tabanlı yukarı yönlü NOMA sezici sisteminde yakın kullanıcıdaki ulaşılabilir hız ve uzak kullanıcıdaki hızlar yukarıda gösterilen Denklem (7) ve (8)'deki gibidir [5].

$$R_1 = \tfrac{1}{2}\log_2(1 + SINR_1) = \tfrac{1}{2}\log_2\left(\frac{P_1|h_1|^2}{P_2|h_2|^2+N_0}\right) \quad (7)$$

$$R_2 = \tfrac{1}{2}\log_2(1 + SINR_2) = \tfrac{1}{2}\log_2\left(\frac{P_2|h_2|^2}{N_0}\right) \quad (8)$$

### III. BENZETİM SONUÇLARI

Bu bölümde çevrimdışı olarak eğitilen yukarı yönlü NOMA ağlarının çevrimiçi sembol sezime işlemi sonrasındaki hata başarımları sunulmuştur. $BS$ ve $UE_i$ arasında ortalama kanal koşulları düşünülerek $UE_1$ ve $UE_2$ için sırasıyla $\sigma_1^2=10$, $\sigma_2^2 = 1$; sayısal modülasyonlar ise QPSK, BPSK seçilmiştir. DL tabanlı kestirimler, literatürdeki kanal bilgisinin bilindiği ML ve SIC-ML sezici ile bulunan sonuçlarla karşılaştırılmıştır. Benzetimlerde Rayleigh sönümlemeli kanal katsayılarının çerçeve boyunca değişmediği, çerçeveden çerçeveye ise değiştiği varsayılmıştır. Benzetim sonuçları SNR (Signal to Noise Ratio- İşaret Gürültü Oranı) değişimine göre verilmiştir. Benzetimlerde bir çerçevede NOMA sembolleri ve pilot sembollerinin sayısı eşittir, bir başka ifade ile pilot/çerçeve oranı ½'dir.

LSTM ağı her bir kullanıcı için ve her bir sezici yöntemi için çevrimdışı ayrı ayrı eğitilerek kaydedilmiştir. Toplamda 10 milyon çerçeve kullanılmıştır. Her bir DL ağı eğitimden sonra kaydedilmiştir. İlgili yöntemler için kaydedilen ağlar çevrimiçi olarak DL tabanlı bit kestirimleri yapılmıştır. DL tabanlı sembol sezim algoritmalarının başarımı, $UE_i$ için gönderilen bit dizileri ile derin öğrenme algoritmalarının kestirdiği bit dizileri arasındaki hatalı bit sayılarının, $UE_i$ için gönderilen toplam bit sayılarına oranlanması ile bulunmaktadır.

Şekil 3'te $UE_1$ için 1) kanal bilgisinin alıcıda bilindiği durumda ML sezici, 2) önerilen iki zaman adımlı DL sezici yöntemi için sırasıyla N=8, 16, 32, 64 ve 128 çerçeve uzunlukları, 3) üç zamanlı DL sezicide N= 8 çerçeve için hata başarımlarını göstermektedir. Şekil 4'te ise $UE_2$'de kanal bilgisinin alıcıda bilindiği durumda SIC-ML sezicinin kullanıldığı durum ile $UE_1$ için verilen aynı yöntemlerle karşılaştırmalı bit hata oranları (bit error rate-BER) verilmiştir.

Şekil 3'ten görüleceği üzere yakın kullanıcının sembolünün tespiti için üç zamanlı DL sezicinin hata başarımları diğer yöntemlerden çok daha iyidir. N=8 için, iki zamanlı DL sezici ile kanal bilgisinin alıcıda bilindiği durumda ML hemen hemen aynı sonuçları vermiştir. $N$ çerçeve uzunluğu arttıkça iki zamanlı DL sezici, ML seziciden daha iyi sonuç vermektedir.

Şekil 4'te, üç zamanlı DL sezicinin SNR 10 dB'den sonra diğer yöntemlerden daha iyi hata performansı verdiği görülmektedir. İki zaman adımlı DL sezici ise N=32 çerçeve uzunluğunda ve çerçeve uzunluğu arttıkça SIC-ML seziciden daha iyi hata performansı sergilemektedir.

Ayrıca benzetimlerden görüldüğü gibi SNR arttıkça klasik ML ve SIC-ML sezicilerde hata olasılık eğrilerinde hata katı (error floor) belirgin şekilde oluşmaktadır. Üç zamanlı DL sezicide SNR arttıkça BER' deki düşüş daha belirgindir. İki zamanlı DL sezicide ise çerçeve sayısı arttıkça hata katı yenilmektedir. Bu durum da yukarı yönlü NOMA'da derin öğrenmenin gücünü perçinlemektedir.

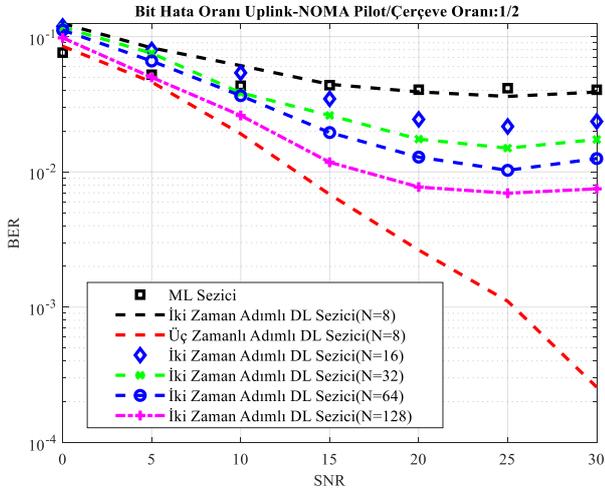

Şekil 3  *UE₁* için Hata Başarımları

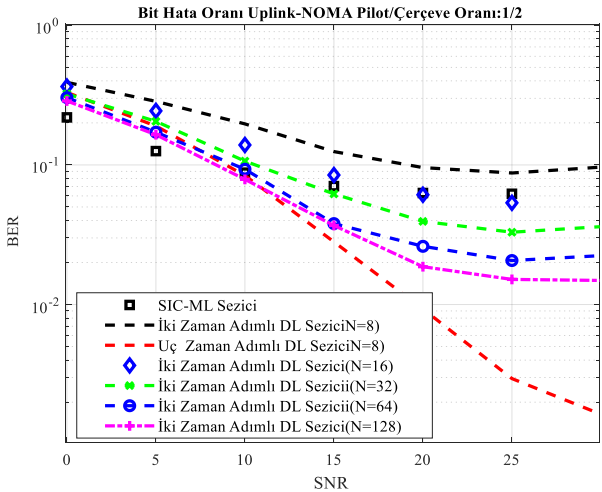

Şekil 4  *UE₂* için Hata Başarımları

## IV. SONUÇLAR VE TARTIŞMA

NOMA sistemlerinde ML ve SIC-ML sezicilere alternatif olarak DL seziciler önerilmiştir. Kanalın alıcıda bilinmediği durumda pilot semboller gönderilerek DL sezicilerin yukarı yönlü haberleşmedeki sembol sezimindeki hata performansları araştırılmıştır. Önerilen DL sezicilerin kanalın bilindiği SIC tabanlı sezicilerden hata performansı açısından üstünlükleri ortaya konmuştur. Üç zaman adımlı DL sezici her durumda ML ve SIC-ML sezicilerden üstündür. İki zaman adımlı DL sezici ise yakın kullanıcı hata performansı açısından ML seziciden üstün olup, uzak kullanıcıdaki hata performansı açısından belirli bir çerçeve sayısından sonra SIC-ML sezicilerden üstün olmaktadır. İki zaman adımlı DL sezici hata performansı açısından üç zaman adımlı DL seziciden iyi olmamasına karşın her iki kullanıcı için tek pilot kullanıldığından zaman kaybını önlemektedir. Bu durum NOMA'nın sağladığı spektral verimin korunmasına yardımcı olmaktadır. Ayrıca yukarı yönlü NOMA sistemlerinde DL sezici kullanımıyla birlikte SNR arttıkça belirgin hale gelen hata olasılık eğrilerindeki hata katı problemi, performans düşüşü, *N* çerçeve boyutunun artmasıyla birlikte azalmaktadır.

Bu çalışma düz sönümleme etkisi gösteren durağımsı (quasistatik) kanal var olduğu kabul edilerek gerçekleştirilmiştir. İlerleyen çalışmalarda hızlı sönümlenmeli sistemlerde benzer etkilerin araştırılması hedeflenmektedir. Dahası gelecekte yapılacak çalışmalarda NOMA sistemlerinde pilot semboller kullanılmadan kör kestirim (blind estimation) temelli DL sezici tasarımı hedeflenmektedir.